\documentclass[12pt]{article}  

\usepackage{cite}
\usepackage{mciteplus}
\usepackage{graphicx}
\usepackage{amssymb}
\usepackage{hyperref}
\usepackage{xspace}
\usepackage{here} 
\usepackage[affil-it]{authblk}

\newcommand{\ud}{\mathrm{d}}
\newcommand{\pt}{\ensuremath{p_{\mathrm{T}}}\xspace}

\begin{document}

\title{An overview of the experimental study of quark-gluon matter in high-energy nucleus-nucleus collisions}

\date{\vspace{-5ex}}

\author{Anton Andronic\thanks{A.Andronic@gsi.de}}

\affil{Reasearch Division and EMMI, GSI Helmholtzzentrum f\"ur Schwerionenforschung, D-64291 Darmstadt, Germany} 

\maketitle

\begin{abstract}
An overview is given on the experimental study of quark-gluon matter produced in 
relativistic nucleus-nucleus collisions, with emphasis on recent measurements at 
the Large Hadron Collider. 

\end{abstract}


\section{Introduction} \label{sect:intro}

The goal of high-energy nucleus-nucleus collisions is to produce and 
characterize a state of nuclear (Quantum Chromo-Dynamics) matter at (energy) 
densities well above the nuclear ground state ($\varepsilon_0\simeq$0.15 GeV/fm$^3$),
an idea now 40 years old \cite{Chapline:1974zf}.
At high densities and/or at high temperatures one expects 
\cite{Collins:1974ky,Cabibbo:1975ig,Chapline:1976gy}
that quarks, the building blocks of hadrons, are no longer confined but move 
freely over distances larger than the size of the nucleon 
(about 1~fm=10$^{-15}$ m).
Such a deconfined state of matter, earlier named the Quark-Gluon 
Plasma (QGP) \cite{Shuryak:1978ij}, was the state of the Universe within 
the first (about 10) microseconds of its creation in the Big Bang 
\cite{Boyanovsky:2006bf} and may exist as well in the core of neutron stars
\cite{Alford:2013pma} (see Ref. \cite{Chapline:1976gy} for an earlier discussion).
The characterization of quark-gluon matter in terms of its equation of state 
(EoS, relating pressure to energy) and of its transport properties (viscosity, 
diffusion coefficients) and delineating its phase diagram \cite{BraunMunzinger:2008tz} 
is a major ongoing research effort \cite{BraunMunzinger:2007zz,Jacak:2012dx,Muller:2013dea,Satz:2013xja,Schukraft:2013wba}. 

At low energies (beam energies per nucleon of up to 10 GeV/$A$ on a fixed 
target, corresponding to center of mass energies per nucleon pair, 
$\sqrt{s_{\mathrm{NN}}}\lesssim$5 GeV) it is expected that compressed nucleonic 
matter is produced (``highly excited nuclear matter'' \cite{Chapline:1974zf}).
The EoS of nuclear matter \cite{Danielewicz:2002pu} at densities a few times
the normal nuclear density ($\rho_0=0.17$ fm$^{-3}=2.7\cdot 10^{14}$ g/cm$^3$), 
expressed as the nuclear compressibility, has relevance for the maximum 
mass of neutron stars (see Ref. \cite{Klahn:2011fb} for an overview).

From lattice QCD calculations, a deconfinement phase transition for an energy 
density of about 1 GeV/fm$^3$ was predicted (see Ref.~\cite{Karsch:2001cy} for 
an early review).
It was shown \cite{Aoki:2006we} that the phase transition at zero 
baryochemical potential, $\mu_B$, is of crossover type, namely with a
continuous, smooth, increase of thermodynamic quantities.
The value of the \mbox{(pseudo-)}critical temperature, $T_c$, at vanishing 
baryochemical potential ($\mu_B$) is currently calculated in lattice 
QCD \cite{Aoki:2009sc,Bazavov:2011nk} to be in the range 155--160 MeV.
The existence of a critical point, denoting the end of the first order 
phase transition line, a point where the phase transition is of a second 
order, is a fundamental question, addressed both experimentally 
\cite{Aggarwal:2010cw} and theoretically \cite{Philipsen:2011zx,Gavai:2014ela}
(see a review in Ref.~\cite{Karsch:2013fga}).

A nucleus-nucleus collision is a highly-dynamical event.
One can identify, schematically, the following stages:
i) initial collisions, occuring during the passage time of the nuclei
($t_{pass}=2R/\gamma_{\mathrm{cm}} c$);
ii) thermalization: equilibrium is established;
iii) expansion and cooling (in a deconfined state);
iv) chemical freeze-out (possibly at hadronization): inelastic collisions 
cease, hadron yields (and distribution over species) are frozen;
v) kinetic freeze-out: elastic collisions cease, spectra and correlations 
are frozen.

The challenge is to characterize the hot (deconfined) stage iii), 
called early-on ``fireball'' \cite{Gosset:1976cy}, while
most of the measurements are performed via hadrons (or their decay products) 
carrying primarily information from the system at stages iv) and v).
Even though the early stage of hot deconfined matter remains inaccessible in a 
direct way because of quark confinement, there are experimental observables 
which carry precious information from this stage.
Extracting the properties of the early hot (deconfined) stage is possible only 
via models.
At low and intermediate energies hadronic transport models 
\cite{Aichelin:1991xy,Danielewicz:2002pu,Petersen:2008kb,Buss:2011mx}
are employed, while at high energies hydrodynamics \cite{Huovinen:2006jp,Gale:2013da} 
is widely used (becoming a ``standard model" for the theoretical understanding
of high energy nucleus-nucleus collisions \cite{Heinz:2013wva}).
Hybrid approaches, combining hydrodynamics and transport, are also employed \cite{Petersen:2008dd}.

Based on model comparison to data in a broad range of collision energies, 
one can extract the following ranges of the fireball characteristics
(values are in the system of units where $\hbar=c=1$ commonly 
used in high-energy physics):
\begin{itemize}
\item {\it Temperature:} $T=100-500$ MeV, or up to a million times the 
temperature at the center of the Sun (\mbox{1 MeV$\simeq$10$^{10}$ K});
\item
{\it Pressure:} $P=100-300$ MeV/fm$^3$ (1 MeV/fm$^3 \simeq 10^{33}$ Pa); 
\item
{\it Density:} $\rho=1-10\cdot\rho_0$;
\item
{\it Volume:} several thousands of fm$^3$; 
\item
{\it Duration:} $10-20$ fm/$c$ (or about $3-6\cdot 10^{-23}$ s).
\end{itemize}

\begin{figure}[hbt]
\centerline{\includegraphics[width=.6\textwidth,height=.53\textwidth]{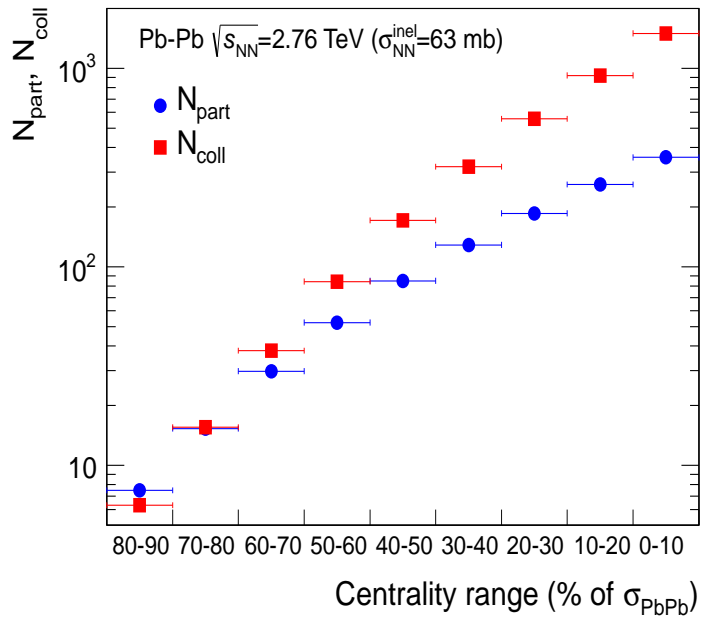}}
\caption{Centrality dependence of $N_{part}$ and $N_{coll}$ (average values) for 
Pb--Pb collisions at $\sqrt{s_{\mathrm{NN}}}=2.76$ TeV (the total interaction cross 
section is $\sigma_{\mathrm{PbPb}}=7.7$ b).
} 
\label{fig:cent} 
\end{figure}

The experimental  ``control parameters'' are:
a) the collision energy (per nucleon pair, $\sqrt{s_{\mathrm{NN}}}$);
b) the centrality of the collision (or, alternatively,
the size of the colliding nuclei). Centrality is obtained from specific 
measurements (see Ref. \cite{Abelev:2013qoq} for such measurements in ALICE).
The usual way of expressing centrality is as percentage of the geometrical 
cross section; another way is via the number of participating 
nucleons, $N_{part}$, namely the nucleons involved in the creation of the 
fireball in the overlap region of the two colliding nuclei \cite{Gosset:1976cy};
$N_{part}$ is calculated as an average over a given centrality range employing 
the Glauber model \cite{Miller:2007ri}. An illustration of the dependence of 
$N_{part}$ and $N_{coll}$ (the number of nucleon-nucleon collisions) on the 
centrality range is presented in Fig.~\ref{fig:cent} for the LHC energy (the 
inelastic nucleon-nucleon cross section, $\sigma_{\mathrm{NN}}^{inel}$, is an input 
for Glauber model calculations; for its energy dependence, see an overview in 
Ref. \cite{Abelev:2012sea}).

After the initial measurements at the Bevalac (Berkeley), 
the program of heavy-ion collisions continued at higher energies at 
Brookhaven at the Alternating Gradient Synchrotron (AGS) and at CERN 
at the Super-Proton Synchrotron (SPS), while in the low energy range 
measurements were performed at GSI Darmstadt at the Schwerionensynchrotron (SIS).
Started in year 2000, the experimental program at the Relativistic Heavy 
Ion Collider (RHIC) at Brookhaven spans $\sqrt{s_{\mathrm{NN}}}$ from 
about 8 to 200 GeV (see, for earlier results, experimental summaries in 
\cite{Arsene:2004fa,Adcox:2004mh,Back:2004je,Adams:2005dq} and an overview
in \cite{Muller:2006ee}).
The study of QCD matter has entered a new era in year 2010 with the 
advent of Pb--Pb collisions at the Large Hadron Collider (LHC), 
delivering the largest ever collision energy, $\sqrt{s_{\mathrm{NN}}}$=2.76 TeV,
more than a factor of 10 larger than previously available. 
An overview of the first LHC data are available in Ref.~\cite{Muller:2012zq,Schukraft:2013wba}.

\section{Global observables} \label{sect:bulk}

\begin{figure}[htb]
\centerline{\includegraphics[width=.62\textwidth,height=.6\textwidth]{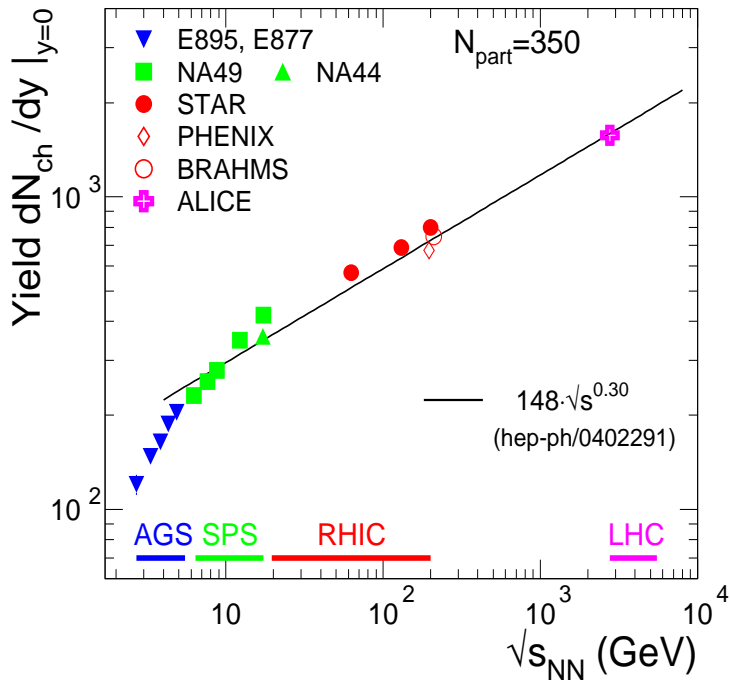}}
\caption{Collision energy dependence of charged-particle rapidity density 
$\ud N_{\mathrm{ch}}/\ud y$  (sum of pions, kaons and protons and their antiparticles) 
at midrapidity, measured by various experiments in central collisions corresponding 
to $N_{part}=350$.
} 
\label{fig:dnchdy} 
\end{figure}

Global (bulk) observables are employed to characterize the bulk thermal
properties of the system. In Fig.~\ref{fig:dnchdy} the collision energy dependence of 
the measured charged-particle rapidity density $\ud N_{\mathrm{ch}}/\ud y$ is shown.
The data are for mid-rapidity, $y$=0 (where particles are emitted in the transverse 
direction); rapidity is defined as:
\begin{equation}
y=\frac{1}{2}\ln\frac{E+p_{\mathrm L}}{E-p_{\mathrm L}}=
\mathrm{tanh}^{-1}(\beta_{\mathrm L}), 
\end{equation}
with $p_{\mathrm L}$ ($\beta_{\mathrm L}$) the longitudinal 
momentum (velocity, in units of $c$), $E=\sqrt{m^2+p^2}$ the total energy.

The measurement of the charged hadrons pseudorapidity 
($\eta=-\ln[\tan(\theta/2)]$, with $\theta$ the polar angle)
density, $\ud N_{\mathrm{ch}}/\ud\eta$, at the LHC was eagerly awaited and 
showed \cite{Aamodt:2010pb} that the increase compared to the measurement at RHIC 
is by a factor of about 2.2 for central collisions. Interpreted as the outcome of 
an increase of the initial entropy density, this increase can be translated into 
a factor of 1.3 increase of the initial temperature \cite{Physics.3.105}. 
The ALICE measurement at the LHC confirmed the phenomenological 
$(\sqrt{s_{\mathrm{NN}}})^{0.3}$ behavior seen at lower energies 
\cite{Andronic:2004tx}, see Fig.~\ref{fig:dnchdy}.
The measurement at the LHC \cite{Aamodt:2010pb} clearly demonstrated that 
the increase of $\ud N_{\mathrm{ch}}/\ud\eta$ with energy is steeper in 
nucleus-nucleus (AA) collisions compared to pp collisions, where the functional 
form is $(\sqrt{s_{\mathrm{NN}}})^{0.22}$.
The data points shown in Fig.~\ref{fig:dnchdy} (left panel) are obtained 
by summing the measured $\ud N/\ud y$ yields for pions, kaons and protons 
and their antiparticles, see below.

\begin{figure}[htb]
\centerline{\includegraphics[width=.64\textwidth,height=.47\textwidth]{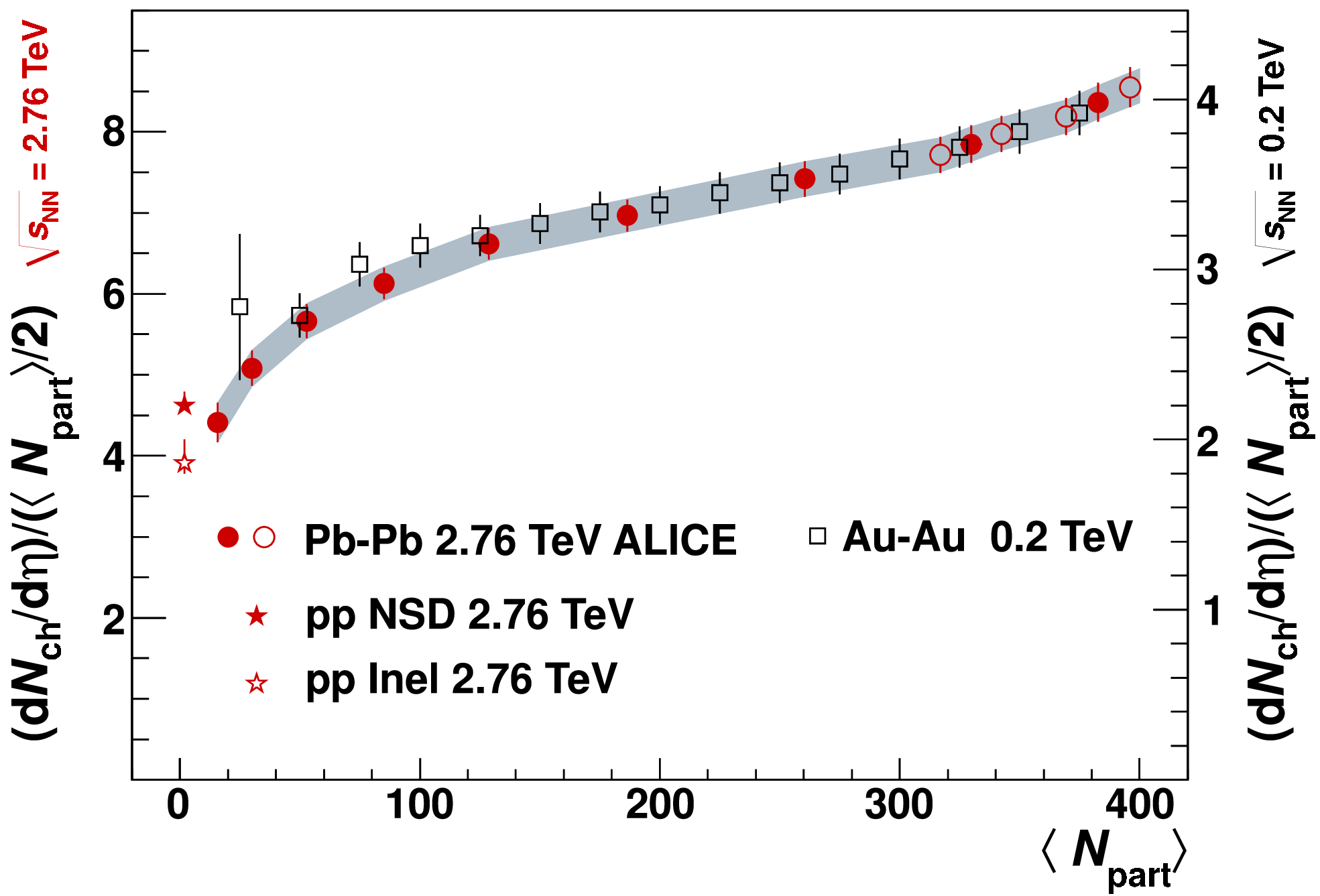}}
\caption{Centrality dependence of the charged particle pseudorapidity density
in Pb--Pb collisions at the LHC and in Au--Au collisions at RHIC (plot from 
Ref. \cite{Aamodt:2010cz}, note the right axis scale for the RHIC data).
} 
\label{fig:dndeta} 
\end{figure}

The centrality dependence of $\ud N_{\mathrm{ch}}/\ud\eta$ is at the LHC 
\cite{Aamodt:2010cz} identical to that measured at RHIC, see Fig.~\ref{fig:dndeta},
pointing to a similar mechanism of particle production at the two energies.
A  model of the parton structure of matter at low parton fractional momentum
$x$, the Color Glass Condensate \cite{Gelis:2010nm,Gelis:2012ri}, describes well the 
data (see an extended comparison to theoretical models in Ref. \cite{Aamodt:2010cz}).

\begin{figure}[htb]
\vspace{-2.9cm}
\centerline{\includegraphics[width=.65\textwidth,height=.8\textwidth]{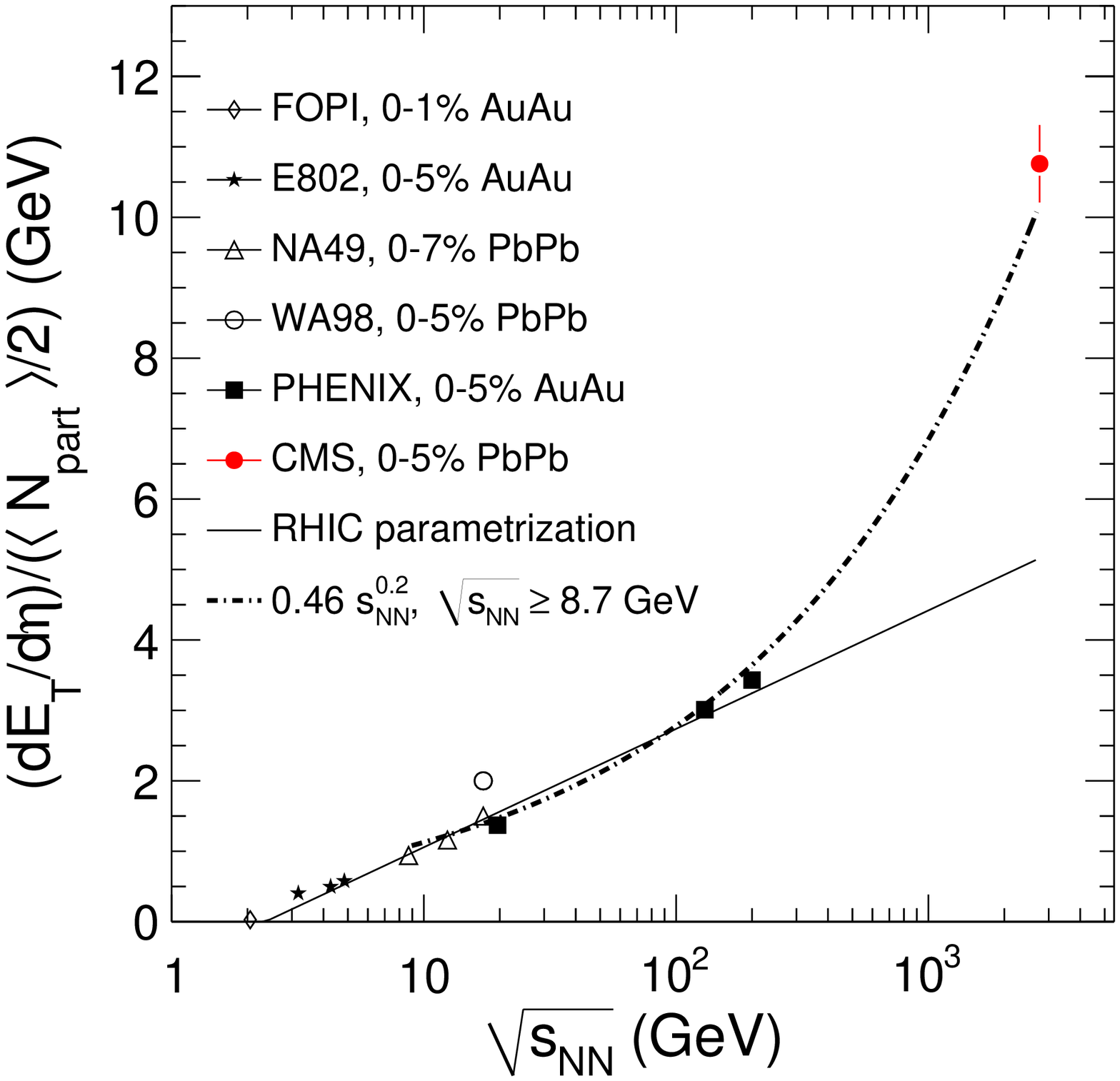}}
\caption{Collision energy dependence of the pseudorapidity density of transverse 
energy in central collisions (plot from Ref. \cite{Chatrchyan:2012mb}).
} 
\label{fig:detdy} 
\end{figure}

Utilizing, in addition to particle counting, the momentum measurement 
(or, alternatively, measuring the total hadron energy in calorimeters),
see Fig.~\ref{fig:detdy}, one can extract from the data the energy density at 
the thermalization time. This involves a space-time model of the collision, which 
was originally put forward by Bjorken \cite{Bjorken:1982qr} and which is presently
incorporated in a more realistic way in hydrodynamical models. 
In the Bjorken model, the energy density is: 
\begin{equation}
\varepsilon = \frac{1}{A_{\rm T}}\frac{\ud E_{\rm T}}{\ud y}\frac{1}{c\tau_0},
\end{equation}
where $E_{\mathrm T}$ is the measured transverse energy and $A_{T}=\pi R^2$ is the
geometric transverse area of the fireball (for central Pb--Pb collisions, 
$A_{T}\simeq$150~fm$^2$).
Assuming a conservative value for the equilibration time, $\tau_0$=1 fm/$c$
(a value which dates back to Bjorken \cite{Bjorken:1982qr}),
one calculates for the LHC energy an energy density of
$\varepsilon_{LHC}$=14 GeV/fm$^3$ \cite{Chatrchyan:2012mb}.
The ``threshold'' value for deconfinement, of about 1 GeV/fm$^3$, from 
lattice QCD calculations \cite{Karsch:2001cy}, is reached at 
$\sqrt{s_{\mathrm{NN}}}\simeq5$ GeV.
Note that the equilibration time may depend on the collision energy; values well
below 1 fm/$c$ are currently used in hydrodynamical models.

\section{Hadron yields and the chemical freeze-out} \label{sect:chem}

\begin{figure}[htb]
\centering\includegraphics[width=.65\textwidth,height=.62\textwidth]{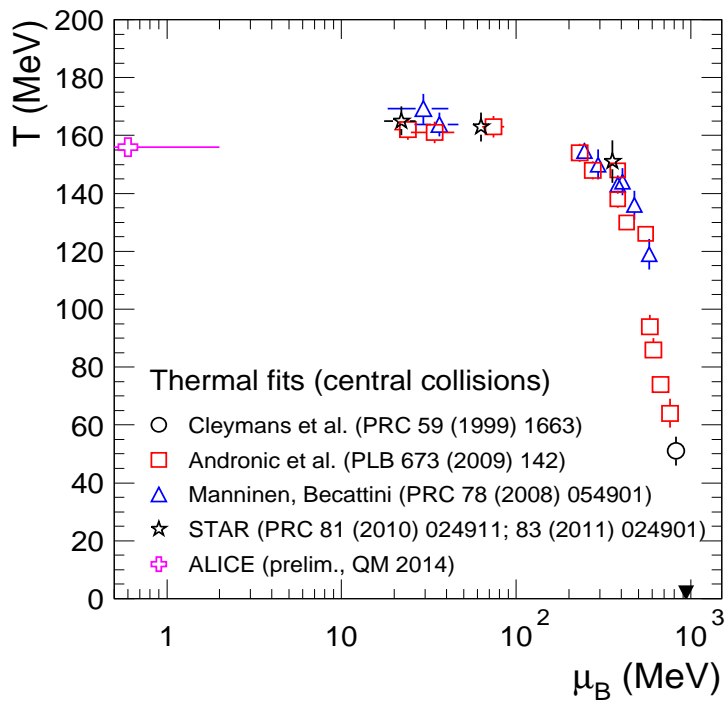}
\caption{The phase diagram of strongly interacting matter with the points
representing the thermal fits of hadron yields at various collision energies
\cite{Cleymans:1998yb,Andronic:2008gu,Manninen:2008mg,Abelev:2009bw,Aggarwal:2010pj,MFloris:2014a}.
For the LHC, $\mu_B$=0 is the outcome of the fit, 0.6 MeV is used here for the sake 
of proper representation with the logarithmic scale.
The down-pointing triangle indicates ground state nuclear matter (atomic 
nuclei).
} 
\label{fig:t-mu}
\end{figure}

In Fig.~\ref{fig:t-mu} the collision energy dependence 
of identified hadron yields at mid-rapidity is shown. 
This comprises measurements, spanning more than 20 years, by experiments at 
the AGS: E895 \cite{Klay:2001tf,Klay:2003zf,Pinkenburg:2001fj}, 
E866/E917 \cite{Ahle:1999uy,Ahle:2000wq}, E891 \cite{Ahmad:1998sg};
the SPS: NA49 \cite{Afanasiev:2002mx,Alt:2007aa,Alt:2005gr,Alt:2008qm}, 
NA44 \cite{Bearden:2002ib},
NA57 \cite{Antinori:2004ee};
RHIC: STAR \cite{Abelev:2008ab,Adler:2002uv,Adams:2006ke,Abelev:2009bw,Aggarwal:2010ig}, 
BRAHMS \cite{Arsene:2005mr}, 
PHENIX \cite{Adler:2003cb};
and the LHC: 
ALICE \cite{Abelev:2012wca,Abelev:2013vea,Abelev:2013xaa,ABELEV:2013zaa,Abelev:2014uua}.

\begin{figure}[htb]
\centering\includegraphics[width=.65\textwidth]{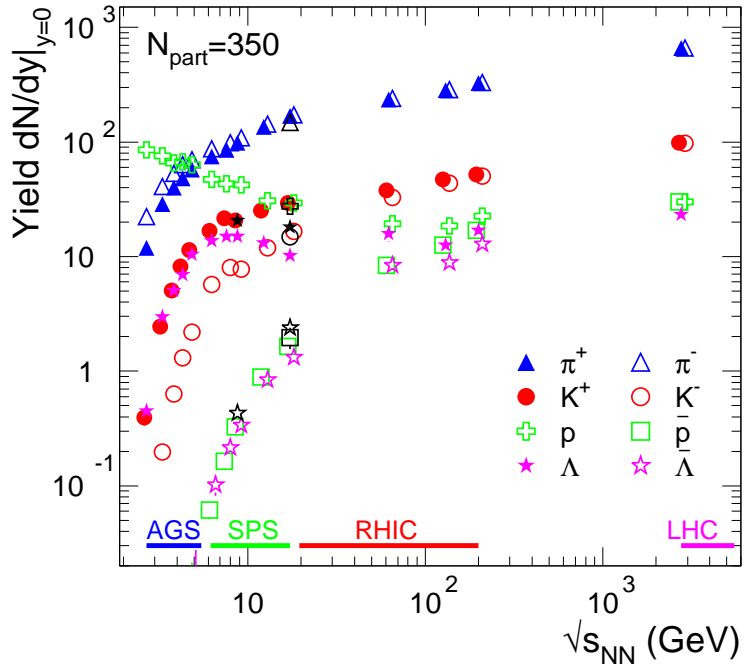}
\caption{Collision energy dependence of the multiplicities (yield, $\ud N/\ud y$, 
at midrapidity) of pions, kaons, protons and $\Lambda$ hyperons and their 
antiparticles, measured in central collisions (corresponding to $N_{part}=350$) of 
Au or Pb nuclei.
} 
\label{fig:dndy}
\end{figure}

The monotonic decrease of the proton yield as a function of energy indicates
that fewer and fewer of the nucleons (or their valence $u,d$ quarks) in the 
colliding nuclei are ``stopped'' in the fireball. An onset of meson production 
is seen, with the kaons (heavier and containing a strange quark) produced 
less abundantly than pions.
The  difference in production yields of $\pi^+$ and $\pi^-$  at low energies 
reflects the isospin composition of the fireball. 
The difference between $K^+$ and $K^-$ meson yields and $\Lambda$ and 
$\bar{\Lambda}$ hyperon production is determined by the quark content
of the hadrons, $K^+$($u\bar{s}$), $K^-$($\bar{u}s$)
$\Lambda$($uds$), $\bar{\Lambda}$($\bar{u}\bar{d}\bar{s}$).  
The availability in the fireball of valence $u$, $d$ 
quarks from colliding nucleons ``stopped" in the fireball leads to a preferential 
production of hadrons carrying those quarks.
These differences vanish gradually for higher energies, where the hadrons are mostly 
newly created (reflecting Einstein's famous equation $m=E/c^2$) and the production 
yields exhibit a clear mass ordering.

\begin{figure}[hbt]
\centering\includegraphics[width=.6\textwidth]{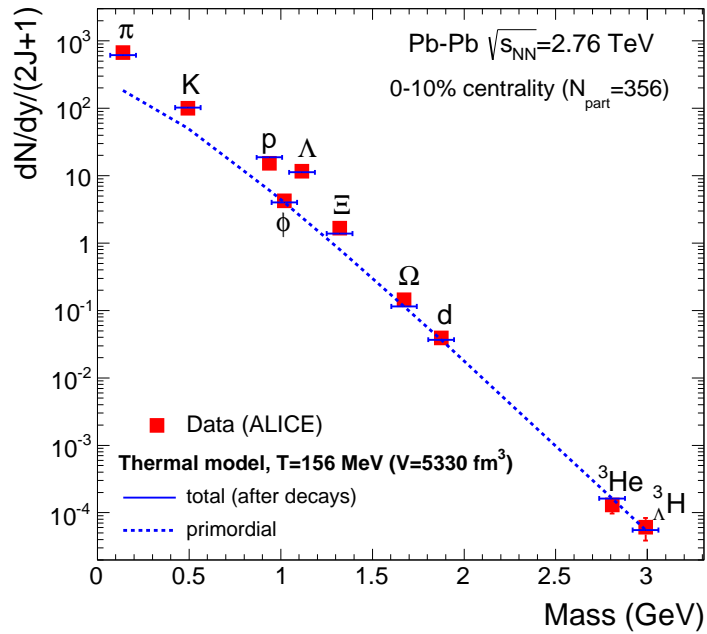}
\caption{Measured hadron abundances divided by the spin degeneracy factor ($2J+1$) 
in comparison with thermal model calculations for the best fit to data 
\cite{MFloris:2014a} in central Pb--Pb collisions at the LHC.
For the model, plotted are the ``total" yields, including all contributions from
high-mass resonances (for the $\Lambda$ hyperon, the contribution from the 
electromagnetic decay $\Sigma^0\rightarrow\Lambda\gamma$, which cannot be resolved 
experimentally, is also included), and the (``primordial") yields prior to decays.
} 
\label{fig:fit}
\end{figure}

Good fits of the measurements are achieved with the thermal model 
\cite{BraunMunzinger:2003zd} with 3 parameters: 
temperature $T$, baryochemical potential $\mu_B$, and volume $V$,
see Fig.~\ref{fig:fit} for the fit of data at the LHC \cite{MFloris:2014a}. 
Remarkably, multiply-strange hyperons and (hyper)nuclei are well
described by the model, which also explains \cite{Andronic:2011yq} the equal 
production of matter and antimatter at the LHC \cite{Abelev:2012wca}.
An interesting question remains whether at hadronization 
the (hyper)nuclei are droplets of quark matter \cite{Chapline:1978kg} or if 
they form via nucleon (and hyperon) coalescence.

The thermal model describes a snapshot of the collision, namely the 
chemical freeze-out, which is assumed to be quasi-instantaneous.
It provides a phenomenological link of data to the QCD phase diagram, 
a link identified early on \cite{Cabibbo:1975ig,Hagedorn:1984hz}
and discussed extensively more recently \cite{Stock:1999hm,BraunMunzinger:2001mh,BraunMunzinger:2003zz,Andronic:2008gu,Andronic:2009gj,Floerchinger:2012xd}.

The phenomenological phase diagram is shown in Fig.~\ref{fig:t-mu}. 
Each point corresponds to a fit of hadron yields in central Au--Au or Pb--Pb 
collisions at a given collision energy. The agreement between the 
results from several independent analyses \cite{Andronic:2008gu,Manninen:2008mg,Abelev:2009bw,Aggarwal:2010pj} is remarkable.
Note that in some cases \cite{Manninen:2008mg,Abelev:2009bw,Aggarwal:2010pj} an
additional fit parameter, the strangeness suppression factor $\gamma_s$, is used
to test possible departure from equilibrium of hadrons containing strange quark(s).
Values of $\gamma_s$ (slightly) below unity are found. 
An approach with more non-equilibrium parameters is also employed 
\cite{Letessier:2005qe,Petran:2013qla}, with somewhat different conclusions.
Fits considering a spread in $T$ and $\mu_B$ were also performed \cite{Dumitru:2005hr}.

A remarkable outcome of these fits is that $T$ increases with increasing 
energy (decreasing $\mu_B$) from about 50 MeV to about 160 MeV, where
it exhibits a saturation for $\mu_B\lesssim$300 MeV.
This saturation of $T$ led to the connection to the QCD phase boundary, via the 
conjecture that the chemical freeze-out temperature can be the hadronization 
temperature \cite{Andronic:2008gu} and that the two regimes in the phase diagram, 
Fig.~\ref{fig:t-mu}, that of approximately constant $T$ for small $\mu_B$ values 
and of the strong increase of $T$ at large $\mu_B$, can imply the existence of a 
triple point in the QCD phase diagram \cite{Andronic:2009gj} 
(see Ref. \cite{Stephanov:1998dy} for an earlier discussion).
Various criteria for the chemical freeze-out were proposed \cite{Tawfik:2004ss,Cleymans:2005xv}.

\begin{figure}[hbt]
\begin{tabular}{lr} \begin{minipage}{.49\textwidth}
\hspace{-0.5cm}\includegraphics[width=1.05\textwidth,height=1.1\textwidth]{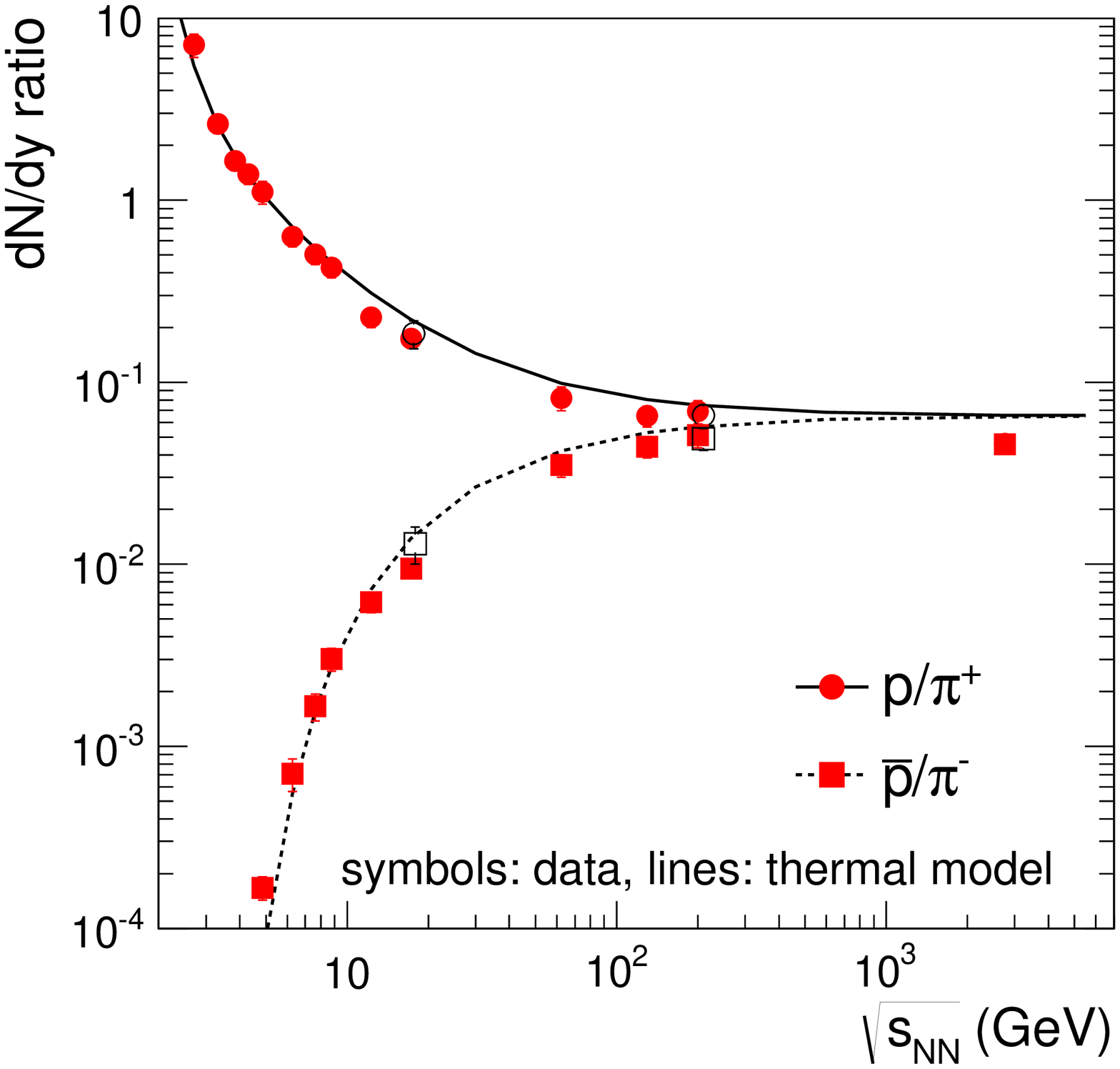}
\end{minipage} & \begin{minipage}{.49\textwidth}
\hspace{-0.5cm}\includegraphics[width=1.05\textwidth,height=1.1\textwidth]{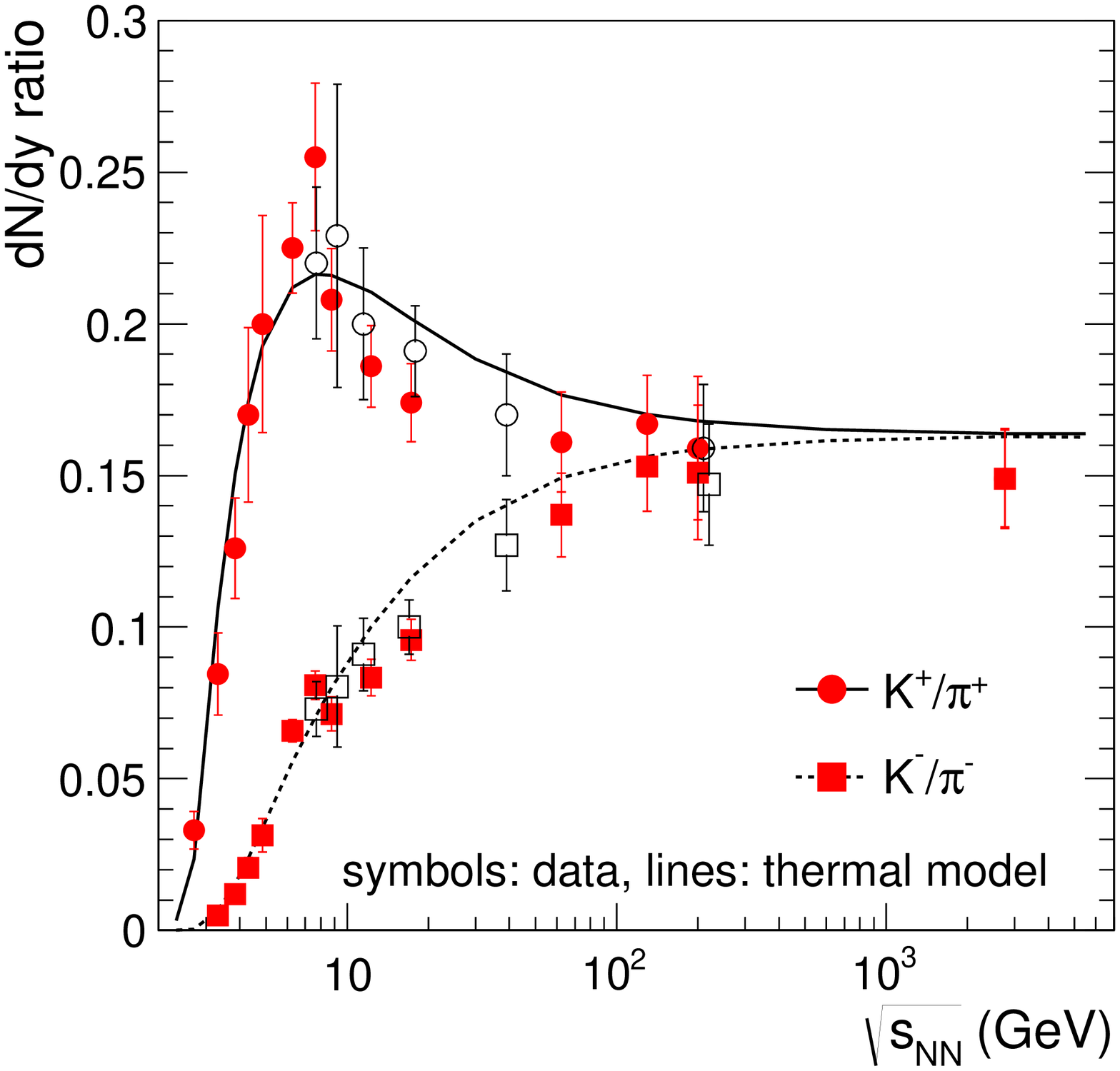}
\end{minipage}\end{tabular}
\caption{Collision energy dependence of ratios of yields of protons and antiprotons 
(left panel) and of kaons (right panel) to yields of pions. The symbols are data, 
the lines are thermal model calculations for energy-dependent parametrizations of 
$T$ and $\mu_B$ (as in Ref. \cite{Andronic:2008gu}).
} 
\label{fig:ratios}
\end{figure}

The thermal model provides the following values of thermodynamical quantities at 
chemical freeze-out (for high energies, corresponding to the ``limiting temperature" 
$T_{lim}=159$ MeV; see Ref. \cite{Andronic:2012ut} for the energy dependence):
pressure $P\simeq 60$ MeV/fm$^3$, energy density $\varepsilon\simeq 330$ MeV/fm$^3$, 
entropy density $s=2.4$ fm$^{-3}$, meson density $n=0.26$ fm$^{-3}$, 
baryon density $n_B=0.06$ fm$^{-3}$ (particles and antiparticles).
In Fig.~\ref{fig:ratios} an illustration is shown of the success of the thermal model 
in reproducing over a broad energy range the ratio of production of various hadron 
species. The calculations are performed with parametrizations
for $T$ and $\mu_B$ as in Ref. \cite{Andronic:2008gu}.
The non-monotonic dependence on energy of the $K^+/\pi^+$ yield ratio was proposed
as a signature \cite{Gazdzicki:1998vd}, and the measurement taken as an
evidence \cite{Alt:2007aa}, for an onset of deconfinement, but can be understood 
well within the hadronic picture of the thermal model \cite{Andronic:2008gu}, 
as shown in Fig.~\ref{fig:ratios} (right panel).
Based on this success, the thermal model predictions provide a reliable guidance
for experimental searches for other exotic nuclei \cite{Andronic:2010qu}.

Current theoretical research addresses the question of flavor-dependent freeze-out 
\cite{Bellwied:2013cta,Bazavov:2013dta};
the role of interactions after chemical freeze-out \cite{Becattini:2014hla}
(in the hybrid model of Ref. \cite{Becattini:2014hla} higher $T$ values are
obtained for the LHC case);
the effect of the extension of hadronic resonance spectrum beyond the currently 
established hadron states \cite{Bazavov:2014xya,Bazavov:2014yba}.
The connection to fits in e$^+$e$^-$ (see Ref. \cite{Andronic:2008ev} and references 
therein) and in elementary hadronic collisions \cite{Becattini:2010sk} remains 
also to be better understood.

Another sector of investigations concerns moments of net-electric charge 
\cite{Adamczyk:2014fia} and net-proton \cite{Adamczyk:2013dal} event-by-event 
multiplicity destributions.
Such measurements have the potential to reveal the critical point in the QCD phase 
diagram \cite{Stephanov:1998dy,Stephanov:1999zu} (see also Ref. \cite{Gavai:2014ela}).
Fluctuations of baryon number and electric charge (determined essentially by the 
light quarks) can be calculated in lattice QCD, allowing an alternative way to 
determine $T$ and $\mu_B$ \cite{Bazavov:2012vg}. Currently, it appears that $T$ values 
extracted in this way \cite{Borsanyi:2014ewa} are somewhat lower that those of the 
fits of hadron yields.

\section{Collective flow and the kinetic freeze-out} \label{sect:kin}

\begin{figure}[hbt]
\centering\includegraphics[width=.6\textwidth]{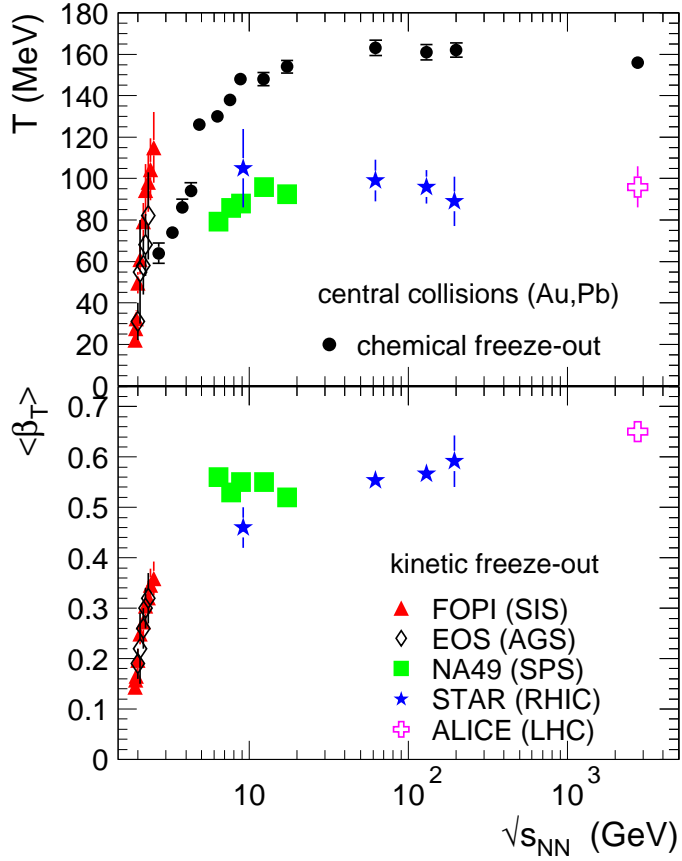}
\caption{Collision energy dependence of collective radial flow in central 
collisions, quantified by the temperature $T$ and average velocity 
$\langle \beta_{\mathrm T}\rangle$ at kinetic freeze-out;
the temperature at chemical freeze-out is represented by the dots.}
\label{fig:flow}
\end{figure}

Collective flow is a distinct feature of nucleus-nucleus collisions, first 
observed in collisions at low energies at the Bevalac \cite{Gustafsson:1984ka}.
In central collisions one investigates the so-called radial flow, which is 
quantified fitting transverse momentum ($p_{\mathrm T}$) spectra with 
the so-called ``blast wave'' model \cite{Schnedermann:1993ws}, obtaining 
in a convenient (albeit simplified) way bulk properties of the 
fireball at kinetic freeze-out.
The extracted fit parameters, the temperature and the average transverse 
velocity $\langle \beta_{\mathrm T}\rangle$, are shown in Fig.~\ref{fig:flow} 
as a function of the collision energy. The measurements are by experiments 
EOS \cite{Lisa:1994yr},
FOPI \cite{Reisdorf:2010aa},
NA49 \cite{Alt:2007uj},
STAR \cite{Abelev:2008ab,Abelev:2009bw},
and ALICE \cite{Abelev:2012wca}.

A strong increase of both $T$ and $\langle\beta_{\mathrm T}\rangle$ is seen 
at low energies (beam energies of up to 1 GeV/$A$ on fixed target).
For $\sqrt{s_{\mathrm{NN}}}\gtrsim5$ GeV, a small further increase of 
$\langle\beta_{\mathrm T}\rangle$ is seen, reaching at the LHC 
$\langle\beta_{\mathrm T}\rangle\simeq$0.65$c$ \cite{Abelev:2012wca}, and a constant 
kinetic freeze-out temperature, which is 50-60 MeV lower than the chemical 
freeze-out value. 
At lower energies, the chemical freeze-out temperature is smaller than the kinetic 
one, which is unphysical and awaits a resolution.

\begin{figure}[hbt]
\centering\includegraphics[width=.63\textwidth]{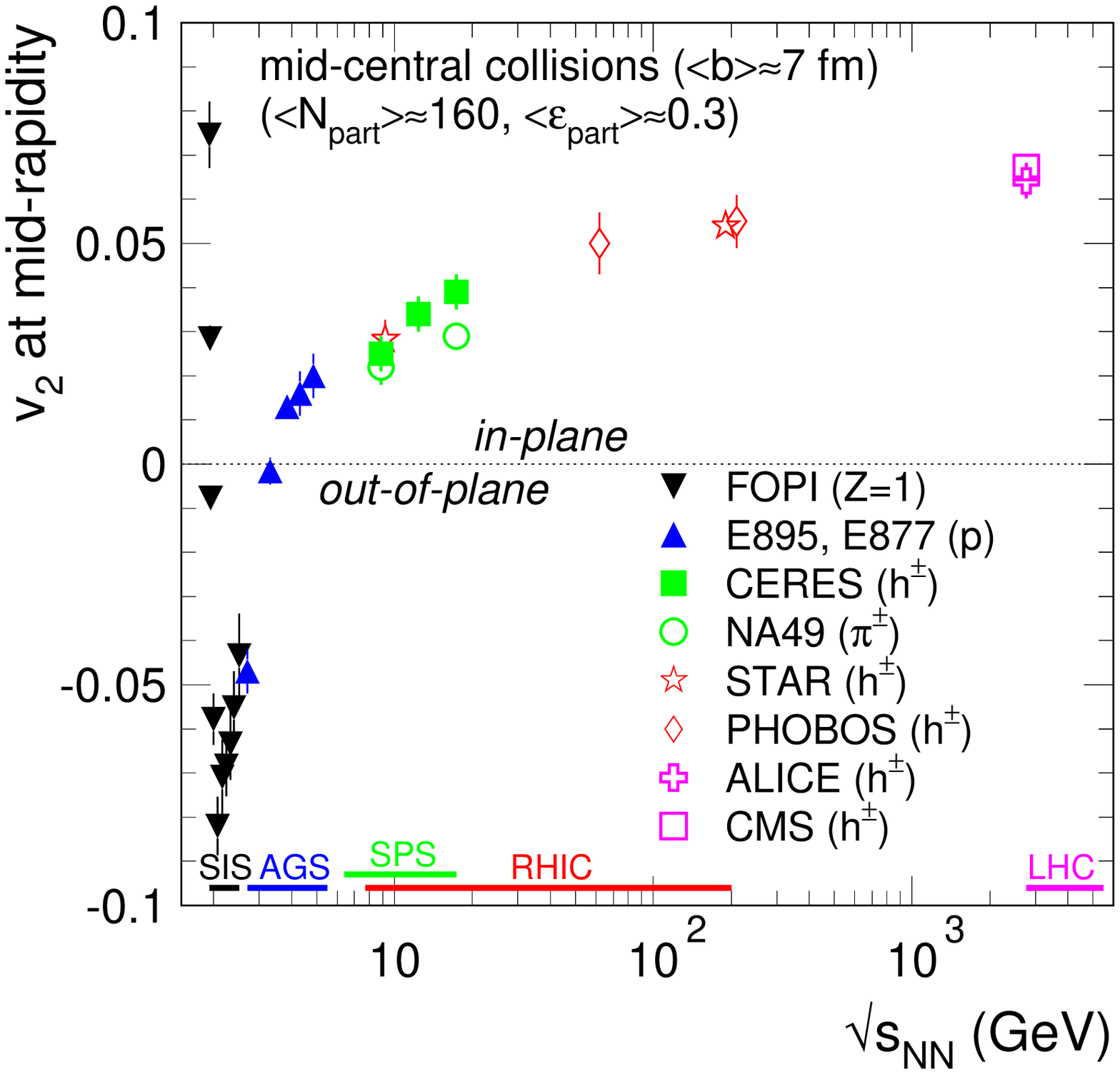}
\caption{Collision energy dependence of the elliptic flow parameter $v_2$, measured 
at midrapidity in mid-central collisions.}
\label{fig:v2}
\end{figure}

In non-central collisions elliptic flow arises as a result of the initial (elliptic)
transverse shape of the overlap zone of the two nuclei (participant eccentricity 
$\varepsilon_{part}$). 
Through the initial gradients of the energy density (or pressure), this
leads to an anisotropic angular emission of hadrons.
This is quantified by the second order (quadrupole) Fourier coefficient
$v_2=\langle \cos(2\varphi)\rangle$, where $\varphi$ is the azimuthal angle with 
respect to the reaction plane.
In Fig.~\ref{fig:v2} we show the energy dependence of elliptic flow,
measured in mid-central collisions ($\langle N_{part}\rangle \simeq 160$, corresponding to an
average impact parameter value of $\langle b\rangle\simeq 7$ fm) by experiments
FOPI \cite{Andronic:2004cp},
E895 \cite{Pinkenburg:1999ya},
E877 \cite{Barrette:1996rs},
CERES \cite{Adamova:2002qx}, NA49 \cite{Alt:2003ab},
STAR \cite{Adams:2004bi}, 
PHOBOS \cite{Alver:2006wh},
ALICE \cite{Aamodt:2010pa}, 
and CMS \cite{Chatrchyan:2012ta}.
The complex evolution of elliptic flow as a function of energy seen
in Fig.~\ref{fig:v2} is understood qualitatively rather well.
At low energies, in-plane ($v_2>0$), rotation-like, emission may arise due
to low energy density in the overlap region and of long reaction times.
The fast transition towards preferential emission out-of-plane ($v_2<0$) 
is the outcome of more energetic collisions, leading to a larger
energy density of the fireball.
The increase of elliptic flow is a fingerprint of a stronger collective 
expansion, hindered by the passing spectators, which act as a shadow for 
the outgoing nucleons and fragments. The competition between the increasing 
speed of the expansion and of the decreasing passage time $t_{pass}$ of 
spectators leads to a maximum of (absolute value) elliptic flow in the SIS 
energy range. In this energy domain, the transiting spectators, with 
$t_{pass}$ varying between 40 and 10 fm/$c$, act as a clock for the collective 
expansion. In this regime, elliptic flow (historically called ``squeeze-out'',
\cite{Demoulins:1990ac,Gutbrod:1988hh})
is a prominent observable for the extraction of the nuclear EoS 
\cite{Danielewicz:2002pu}.
Towards larger energies, elliptic flow exhibits another transition, to
a preferential in-plane emission \cite{Ollitrault:1992bk}, a result of 
a unhindered collective expansion of the initially-anisotropic fireball.
Elliptic flow is built mostly in the earlier stages of the collision, 
since it is determined by the initial pressure gradients, 
which it alters quickly as it develops.
Consequently, at high energies, elliptic flow probes (albeit not 
exclusively) the deconfined state of quarks and gluons.

The elliptic flow measurement at the LHC \cite{Aamodt:2010pa} 
exhibits a magnitude about 30\% larger compared to the measurement at 
$\sqrt{s_{\mathrm{NN}}}=200$ GeV. This increase is described by hydrodynamics 
and was anticipated on a purely phenomenological $\log(\sqrt{s_{\mathrm{NN}}})$ 
behavior seen at lower energies \cite{Andronic:2004tx}.
The data show  \cite{Aamodt:2010pa} that the $p_{\mathrm{T}}$ dependence of $v_2$ is 
identical at the LHC to that measured at RHIC, implying that the increase for the 
$p_{\mathrm{T}}$-integrated $v_2$ value is due exclusively to the increase of the 
average transverse momentum of the hadrons. 
The scaling of $v_2$ with the number of valence quarks observed at RHIC 
\cite{Adler:2003kt,Adams:2003am} seems less well obeyed at the 
LHC \cite{Abelev:2014pua}.

\begin{figure}[hbt]
\centering\includegraphics[width=.9\textwidth]{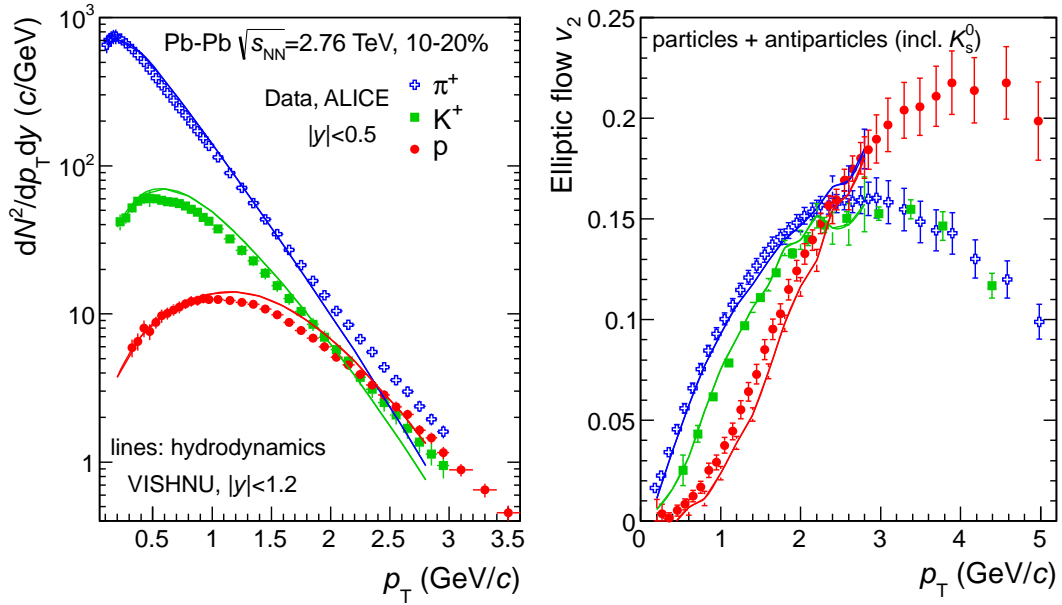}
\caption{Transverse momentum spectra and the transverse momentum dependence of 
the elliptic flow parameter $v_2$ for pions, kaons and protons measured in
10-20\% central Pb--Pb collisions at $\sqrt{s_{\mathrm{NN}}}=2.76$ TeV at the LHC 
\cite{Abelev:2013vea,Abelev:2014pua}, in comparison to hydrodynamical model calculations
\cite{Song:2013qma}.}
\label{fig:v2pt}
\end{figure}

A good description of spectra and elliptic flow measured at the LHC is achieved in 
hydrodynamical models, both at RHIC and at the LHC energies, see 
Ref. \cite{Abelev:2013vea,Abelev:2014pua} for the LHC case; 
an illustration for the VISHNU model \cite{Song:2013qma} is shown in 
Fig.~\ref{fig:v2pt}.
Another component of flow, called directed flow, is quantified by the first Fourier 
coefficient, $v_1=\langle \cos(\phi)\rangle$. 
Directed flow is strong at lower energies, where it was observed observed for the
first time \cite{Gustafsson:1984ka},
and is studied currently both at the LHC \cite{Abelev:2013cva} and at RHIC 
\cite{Adamczyk:2014bea} energies.
Higher-order Fourier coefficients are also studied extensively \cite{ALICE:2011ab,Aad:2013xma}, while further refinements based on multi-particle azimuthal correlations 
\cite{Abelev:2014mda} address the question of collective phenomena in small systems.
Flow-like features have been indeed identified recently in p--Pb collisions at the LHC
\cite{Abelev:2012ola,Aad:2012gla,Chatrchyan:2013nka,ABELEV:2013wsa} and at 
RHIC \cite{Adare:2013piz};
some of these features were earlier predicted by hydrodynamical models 
\cite{Bozek:2012gr}, incorporating a collective expansion akin to that in Pb--Pb 
collisions.
This is currently a subject of intense theoretical effort \cite{Dusling:2012iga,Dusling:2013oia,Werner:2013tya,Qin:2013bha,Bzdak:2013zma,Bozek:2013ska}, 
that promises to bring further insights on flow effects in small volumes 
(but characterized by very large energy density).

Another set of experimental observables from the kinetic freeze-out stage are
obtained from Bose-Einstein correlations of identical particles (called also 
femtoscopy, or Hanbury Brown and Twiss (HBT) interferometry) \cite{Lisa:2005dd}.
This enables the measurement of the spatial extension of a region of the fireball, 
the region of homegeneity, and the lifetime of the expanding fireball until the 
kinetic freeze-out.
These measurements, performed over twenty years of heavy-ion collision studies, 
have recently been extended with data at the LHC \cite{Aamodt:2011mr} and at 
RHIC \cite{Adamczyk:2014mxp}.

\begin{figure}[hbt]
\centering\includegraphics[width=.61\textwidth]{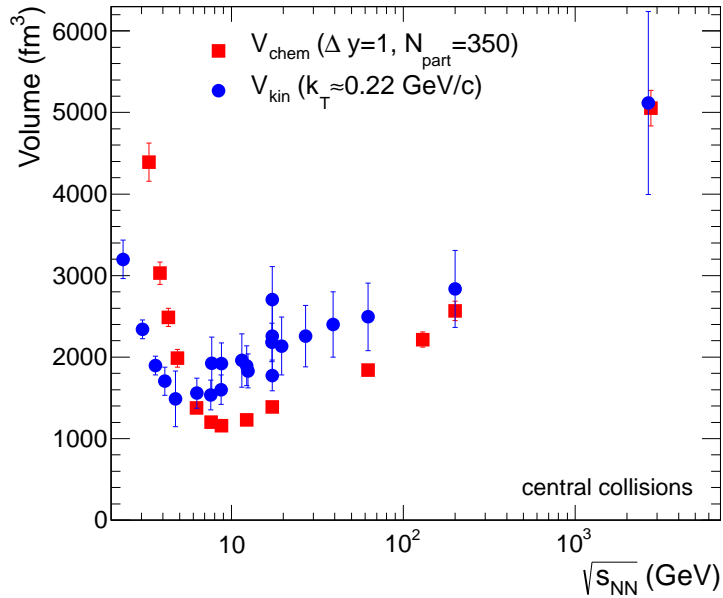}
\caption{Collision energy dependence of the fireball volume extracted at chemical 
and at kinetic freeze-out in central collisions. The data for the kinetic freeze-out
volume is from Ref. \cite{Adamczyk:2014mxp}.}
\label{fig:vol}
\end{figure}

The energy dependence of the volume extracted at chemical freeze-out 
$V_{chem}$ and at kinetic freeze-out $V_{kin}$ in central collisions is shown
in Fig.~\ref{fig:vol}.
$V_{chem}$ corresponds to one unit of rapidity and is calculated from the 
$\ud N_{ch}/\ud y$ data (Fig.~\ref{fig:dnchdy}) and the densities calculated in 
the thermal model for parametrized $T$ and $\mu_B$ values.
$V_{kin}$ is obtained from femtoscopy radii and corresponds to the pion pair 
transverse momentum $k_{\mathrm T}\simeq 0.22$ GeV/$c$ 
(data from Ref. \cite{Adamczyk:2014mxp}, see references therein).
Note that the kinetic freeze-out volume extracted from femtoscopy is not the entire
volume of the fireball, but that of the region of homogeneity \cite{Lisa:2005dd}.

\begin{figure}[hbt]
\centering\includegraphics[width=.58\textwidth]{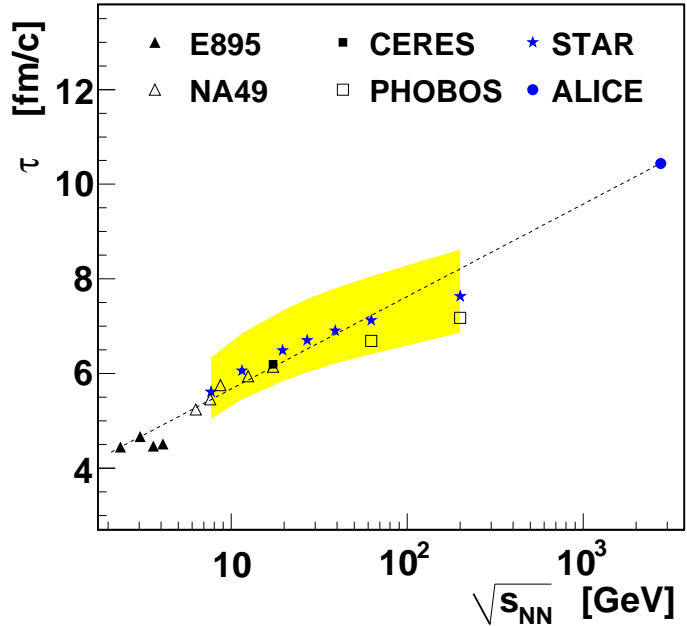}
\caption{Collision energy dependence of the duration until kinetic freeze-out, $\tau$,
for central collisions (plot from Ref. \cite{Adamczyk:2014mxp}, see references 
therein).
The yellow band denotes the uncertainty in $\tau$ due to the uncertainty in the 
freeze-out temperature \cite{Adamczyk:2014mxp}.
}
\label{fig:tau}
\end{figure}

The energy dependence of the (``decoupling") time at kinetic freeze-out 
$\tau$, obtained from fits of $R_{long}(k_{\mathrm T})$ assuming a one-dimensional 
(longitudinal) hydrodynamics expansion (the Bjorken model), is shown in 
Fig.~\ref{fig:tau}.
Hydrodynamical models \cite{Karpenko:2012yf,Bozek:2014xsa}, are successful in 
predicting the femtoscopy observables (see Ref. \cite{Aamodt:2011mr}).

Elliptic flow, radial flow and femtoscopy observables are used to extract, via 
comparisons to hydrodynamic calculations, the ratio of the shear viscosity to 
entropy density, $\eta/s$ \cite{Physics.2.88}).
The remarkable description of flow and femtoscopy in hydrodynamic 
models observed at RHIC is further confirmed and extended with the 
measurements at the LHC.
The quantitative determination of $\eta/s$ is dependent on fine details of the 
initial energy density distribution \cite{Song:2010mg}, which can be calculated, 
based on constraints on average values from data, according to either the Glauber 
model or in the Color Glass Condensate framework; 
recent work \cite{Retinskaya:2013gca} aims at quantitative constraints for the 
initial conditions. 
Deconfined matter is characterized by low values of $\eta/s$ 
\cite{Luzum:2008cw,Song:2010mg,Gardim:2012yp} and it remains a challenge to
establish the possible existence of a lower bound \cite{Kovtun:2004de};
for values extracted at much lower collision energies see Ref. \cite{Danielewicz:1984kt,Schmidt:1993ak};
an estimate for nuclear matter in mildly-excited nuclei is given in
Ref.~\cite{Auerbach:2009ba}.
Thermalization at a very early stage of the collision, at or below 1 fm/$c$, 
as employed in hydrodynamic calculations, remains a challenge to 
theory \cite{Berges:2012ks}, but recent developments \cite{Kurkela:2014tea}
bring further hope for understanding the issue.

Can we identify an onset of deconfinement based on the (bulk) hadronic observables
discussed above?  
The change, in the range $\sqrt{s_{\mathrm{NN}}}\simeq$5-10 GeV, of fireball
properties (see Fig.~\ref{fig:flow}, \ref{fig:v2}) is a possible fingerprint,
but further experimental and theoretical support is needed to conclude.

Probing the deconfined matter in a more direct way is done with special
observables of the early stage. 
Proposed early on as thermometers of the deconfined stage, thermal photons and 
low-mass dileptons \cite{Kajantie:1981wg} (see Ref. \cite{Gale:2012xq,Rapp:2013nxa}
for recent reviews) are such observables.
Low-mass dileptons are also probes of the chiral symmetry restoration in 
hot QCD matter \cite{Rapp:2009yu}.
Measurements of thermal photons at RHIC have shown \cite{Adare:2008ab} 
(see also Ref. \cite{Physics.3.28}) that the temperature averaged over the lifetime 
of the fireball is larger than the chemical freeze-out $T$.
A somewhat larger $T$ value is extracted from preliminary data of photon production 
at the LHC \cite{Wilde:2012wc}.
The observation of elliptic flow of thermal photons at RHIC \cite{Adare:2011zr},
observed also at the LHC \cite{Lohner:2012ct}, made more complex the connection 
between the thermal photon measurements and the hottest stage of the system
\cite{vanHees:2011vb,Shen:2013vja}. This is a challenging and fascinating field 
of investigation, both experimentally and theoretically, from which crucial insights
are expected to arise in the coming years.

Another category of QGP probes is the so-called hard probes, namely 
processes characterized by an energy scale (quantified by the transverse mass
$m_{\mathrm T}=\sqrt{m_0^2+p_{\mathrm T}^2}$, where $m_0$ is the rest mass of the 
hadron) above several GeV (well above the temperature of the medium). 
Examples of such observables are hadrons at high $p_{\mathrm T}$ (or jets) 
\cite{d'Enterria:2009am} and hadrons containing heavy (charm or bottom) 
quarks \cite{Averbeck:2013oga}; they are produced at early times in the collision, 
$t=1/m_{\mathrm T}$.
Quarkonium formation \cite{Matsui:1986dk} is another prominent 
observable for deconfined matter studies; 
sections \ref{sect:eloss} and \ref{sect:quarkonium} present selected results on
these measurements and their interpretation in theoretical models.

\section{Parton energy loss} \label{sect:eloss}

Proposed by Bjorken in 1982 \cite{Bjorken:1982tu}, ``jet quenching'',
the extinction of jets (due to the energy loss of the parent parton) in QGP 
was for the first time observed at RHIC \cite{Adcox:2001jp,Adler:2002xw} and is a 
subject of intense study at the LHC 
\cite{Aamodt:2010jd,CMS:2012aa,Abelev:2012eq,Chatrchyan:2012gw,Aad:2012vca}.
The usual method to quantify jet quenching is via the nuclear modification 
factor, defined as: 
\begin{equation}
R_{\mathrm{AA}}=\frac{\ud^2 N_{{\mathrm AA}}/\ud y\ud p_{\mathrm T}}{N_{coll}\cdot\ud^2 N_{\mathrm{pp}}/\ud y\ud p_{\mathrm T}}, 
\end{equation}
where $\ud^2 N/\ud y\ud p_{\mathrm T}$ denotes the yield of a given observable measured in 
nucleus--nucleus (AA) or proton--proton (pp) collisions and $N_{coll}$ is 
the average number of nucleon-nucleon collisions over the given centrality 
interval of AA collisions (see Fig.~\ref{fig:cent}).

A change of physics in AA collisions (which in specialized terms is called 
a ``medium effect'') is seen as a departure of $R_{\mathrm{AA}}$ from unity.
Modifications of parton distributions in nuclei compared to free nucleons,
denoted as ``shadowing" or ``saturation", need to be considered carefully, in 
particular at LHC energies.
Measurements in p--Pb collisions address this issue \cite{ALICE:2012mj}.

We note that the binary collision scaling assumed in the construction of 
$R_{\mathrm{AA}}$ applies only to hard processes.
It is known experimentally that bulk particle production (comprising 
essentially pions, protons and kaons at low-momentum, 
$p_{\mathrm T}\lesssim$3-4 GeV/$c$) in AA collisions scales (in first order) 
with $N_{part}$ \cite{Aamodt:2010cz} (see Fig.~\ref{fig:dndeta}).

\begin{figure}[hbt]
\centerline{\includegraphics[width=.67\textwidth]{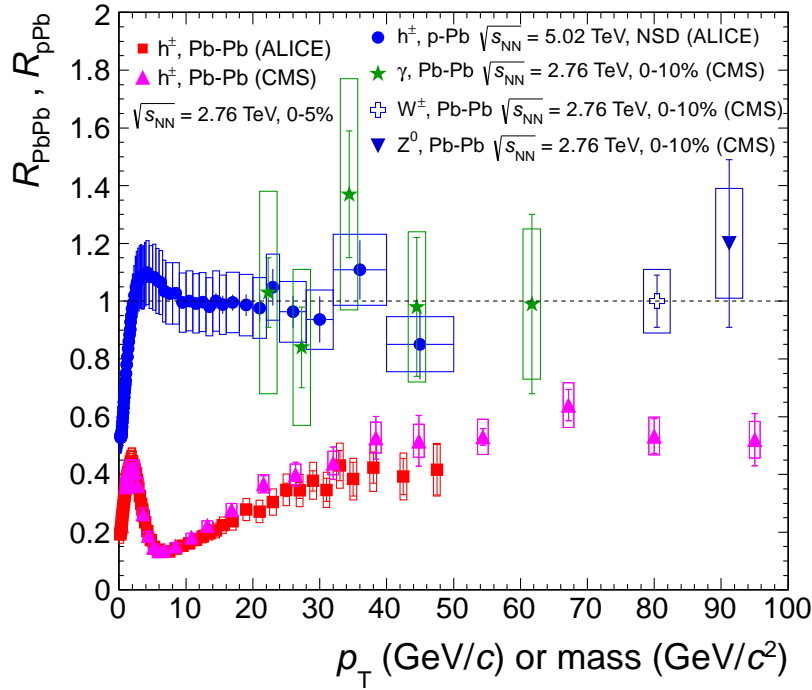}}
\caption{Transverse momentum dependence of the nuclear modification 
factor $R_{\mathrm{pPb}}$ of charged particles (h$^\pm$) measured in minimum-bias (NSD) 
\mbox{p--Pb} collisions at $\sqrt{s_{\mathrm{NN}}} = 5.02$~TeV \cite{Abelev:2014dsa}
in comparison to data on the nuclear modification factor $R_{\mathrm{PbPb}}$ in central 
Pb--Pb collisions at $\sqrt{s_{\mathrm{NN}}} = 2.76$~TeV.
The Pb--Pb data are for charged particle \cite{CMS:2012aa,Abelev:2012eq},
direct photon \cite{Chatrchyan:2012vq}, Z$^0$ \cite{Chatrchyan:2011ua},
and W$^{\pm}$ \cite{Chatrchyan:2012nt} production. All data are for midrapidity
(plot from Ref. \cite{Abelev:2014dsa}).}
\label{fig:raa_all} 
\end{figure}

The first measurement of $R_{\mathrm{AA}}$ for inclusive charged-hadron production at 
the LHC \cite{Aamodt:2010jd} showed that the suppression is larger than 
previously measured at RHIC, reaching a factor of about 7. 
The suppression is reduced towards larger $p_{\mathrm T}$ values, 
see Fig.~\ref{fig:raa_all}, but remains substantial even at 
50-100 GeV/$c$ \cite{CMS:2012aa,Abelev:2012eq}.
Recent data for p--Pb collisions at the LHC \cite{Abelev:2014dsa} demonstrate that 
the suppression of hadron production at high \pt in Pb--Pb collisions, has no 
contribution from initial state effects.
The ALICE p--Pb data show no sign of nuclear matter modification of
hadron production at high \pt and are therefore fully consistent with the 
observation of binary  collision scaling in Pb--Pb of observables which are not 
affected by hot QCD matter (direct photons \cite{Chatrchyan:2012vq} and vector 
bosons \cite{Chatrchyan:2011ua,Chatrchyan:2012nt}).

\begin{figure}[hbt]
\centerline{\includegraphics[width=.67\textwidth]{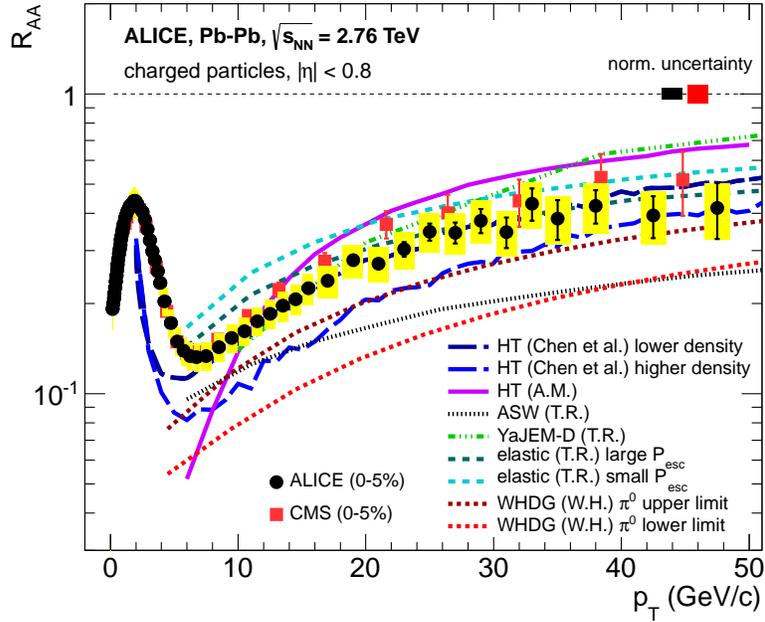}}
\caption{The nuclear modification factor $R_{\mathrm{AA}}$ as a 
function of transverse momentum for inclusive charged hadron production in Pb--Pb 
collisions at the LHC (Fig. from Ref. \cite{Abelev:2012eq}, see references therein 
for the theoretical curves).}
\label{fig:raa_theo} 
\end{figure}

The measurements, in conjunction with theoretical models, clearly demonstrate
that partons lose energy in the deconfined hot and dense matter created in collisions
at RHIC and LHC.
The basic features seen in the data are reproduced by models,
see Fig.~\ref{fig:raa_theo} for the LHC energy. 
The ultimate goal of such studies is the extraction of transport coefficients; 
presently, large uncertainties are originating from the description of jet quenching 
in theoretical models, which remains a challenging task \cite{Renk:2012wi}.
The spread of the theoretical curves in Fig.~\ref{fig:raa_theo} illustrates this 
challenge.
Recent progress on the theoretical front and the availability of data at the LHC 
allows already a quantitative extraction of the jet quenching parameter 
$\hat{q}$ (which is the average squared transverse momentum 
acquired by the parton per unit path length) in deconfined QCD 
matter, a quantity which has recently been calculated 
in lattice QCD \cite{Panero:2013pla}.
Values of $\hat{q}$ of several GeV$^2$/fm are extracted in the systematic study of
Ref. \cite{Burke:2013yra}.

\begin{figure}[hbt]
\centerline{\includegraphics[width=.62\textwidth,height=.58\textwidth]{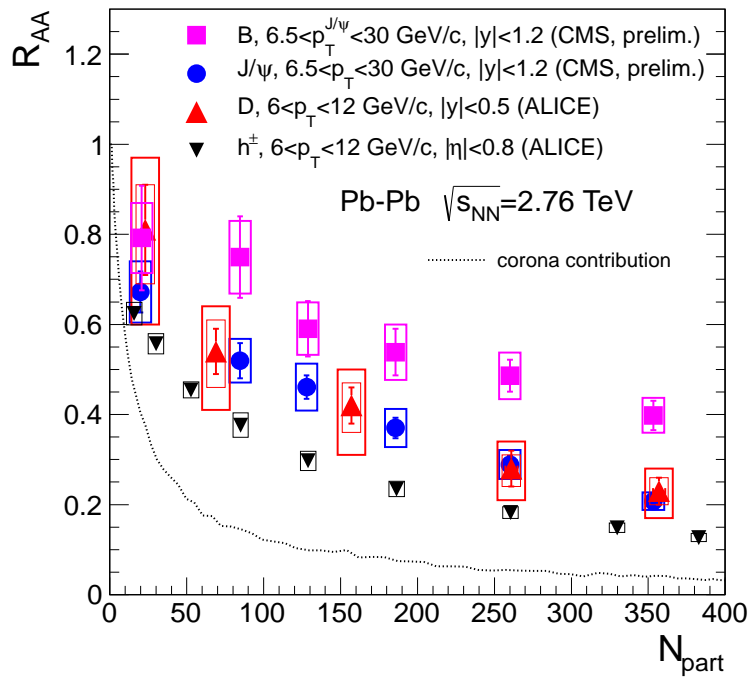}}
\caption{Centrality dependence of the nuclear modification factor $R_{\mathrm{AA}}$ 
for charged hadrons at high $p_{\mathrm T}$ in comparison to that of charmed mesons ($D$),
charmonium (J/$\psi$) and beauty hadrons ($B$). 
The dotted line is an estimate of the lowest $R_{\mathrm{AA}}$ value, originating from
the production in the corona of colliding nuclei.
} 
\label{fig:raa_comp} 
\end{figure}

The nuclear modification factor for production of hadrons carrying charm 
\cite{ALICE:2012ab} or bottom \cite{CMS-PAS-HIN-12-014} quarks 
is shown  in Fig.~\ref{fig:raa_comp} as a function of centrality in Pb--Pb collisions
at the LHC. 
The theoretical expectation is that heavy quarks (charm and bottom) lose 
less energy (by gluon radiation) compared  to lighter (up, down, strange) 
ones \cite{Dokshitzer:2001zm}. 
This expectation appears to be exhibited by the data, see Fig.~\ref{fig:raa_comp},
although a definite conclusion needs further support from experiment.

The energy loss suffered by energetic heavy quarks in QGP 
is indicative of their ``strong coupling'' with the medium, dominated by 
light quarks and gluons.
The measurements at the LHC \cite{Adare:2010de} consolidate and extend the 
observation at RHIC of heavy quark energy loss and flow \cite{Adare:2006nq}.
Theoretical models of quark energy loss in deconfined matter describe the data
well, although a large spread exists between various model predictions
(see Ref. \cite{Adare:2010de} and references therein).
New data at RHIC with reconstructed $D^0$ mesons\cite{Adamczyk:2014uip} indicate
similar mechanisms of energy loss compared to the LHC case.
Measurements of elliptic flow of heavy quarks at the LHC \cite{Abelev:2014ipa} and 
at RHIC \cite{Adare:2006nq,Adamczyk:2014yew} impose  additional constraints to
the theoretical models \cite{He:2014cla}.

Experimental studies at the LHC with reconstructed jets \cite{Aad:2012vca,Chatrchyan:2013kwa,Abelev:2013kqa} indicate a redistribution of the energy inside the jet 
cone \cite{Chatrchyan:2013kwa}.
Measurements of jets with bottom quarks indicate that parton energy loss at higher 
parton energies ($80-250$ GeV) is not dependent on parton mass (flavor)
\cite{Chatrchyan:2013exa}.

\section{Quarkonium} \label{sect:quarkonium}

Among the various suggested probes of deconfinement, charmonium ($c\bar{c}$) 
states plays a distinctive role. The J/$\psi$ meson is the first hadron for which 
a clear mechanism of suppression (melting) in QGP was proposed early on, 
based on the color analogue of Debye screening \cite{Matsui:1986dk}.
Further refinements, including the whole quarkonia species, $c\bar{c}$ and
$b\bar{b}$, led to the picture of ``sequential suppression" 
\cite{Karsch:1990wi,Digal:2001ue,Karsch:2005nk},
a hierarchy of quarkonium dissociation dependent on the binding energy (size) 
of the quarkonium state, which could give information on the temperature of the 
medium (given that the Debye length in deconfined matter has a pronunced 
temperature dependence \cite{Digal:2001ue}).
It was pointed out early-on that the Debye screening phenomenon is a low-$p_{\mathrm T}$ 
effect \cite{Karsch:1990wi}. 
A review of data and its interpretation in the screening scenario is available in
Ref. \cite{Kluberg:2009wc}.
Lattice QCD calculations can give information on the screening, see earlier
arguments \cite{Asakawa:2003re,Mocsy:2007yj,Mocsy:2007jz} and more recent work 
\cite{Morita:2010pd,Ding:2012sp,Lee:2013dca,Hayata:2012rw,Rothkopf:2013kya,Borsanyi:2014vka} (see a review in Ref. \cite{Mocsy:2013syh}).

Novel quarkonium production mechanisms were proposed.
In the statistical hadronization model \cite{BraunMunzinger:2000px}, the 
charm quarks produced in initial hard collisions thermalize in QGP and 
are ``distributed'' into hadrons at chemical freeze-out. 
All charmonium states are assumed to be not formed at all in the deconfined 
state but are produced, together with all other hadrons, at chemical 
freeze-out \cite{Andronic:2006ky,BraunMunzinger:2000px} 
(see Ref. \cite{BraunMunzinger:2009ih,Andronic:2011yq} for recent predictions of 
this model).
Kinetic (re)combination of charm and anti-charm quarks in QGP \cite{Thews:2000rj}
is an alternative quarkonium production mechanism.
In this model (see Ref. \cite{Liu:2009nb,Zhao:2011cv,Emerick:2011xu,Zhou:2014kka} 
for recent results), continuous dissociation and (re)generation of charmonium takes 
place over the entire lifetime of the deconfined stage. 

\begin{figure}[hbt]
\begin{tabular}{lr} \begin{minipage}{.49\textwidth}
\hspace{-0.3cm}\includegraphics[width=1.02\textwidth]{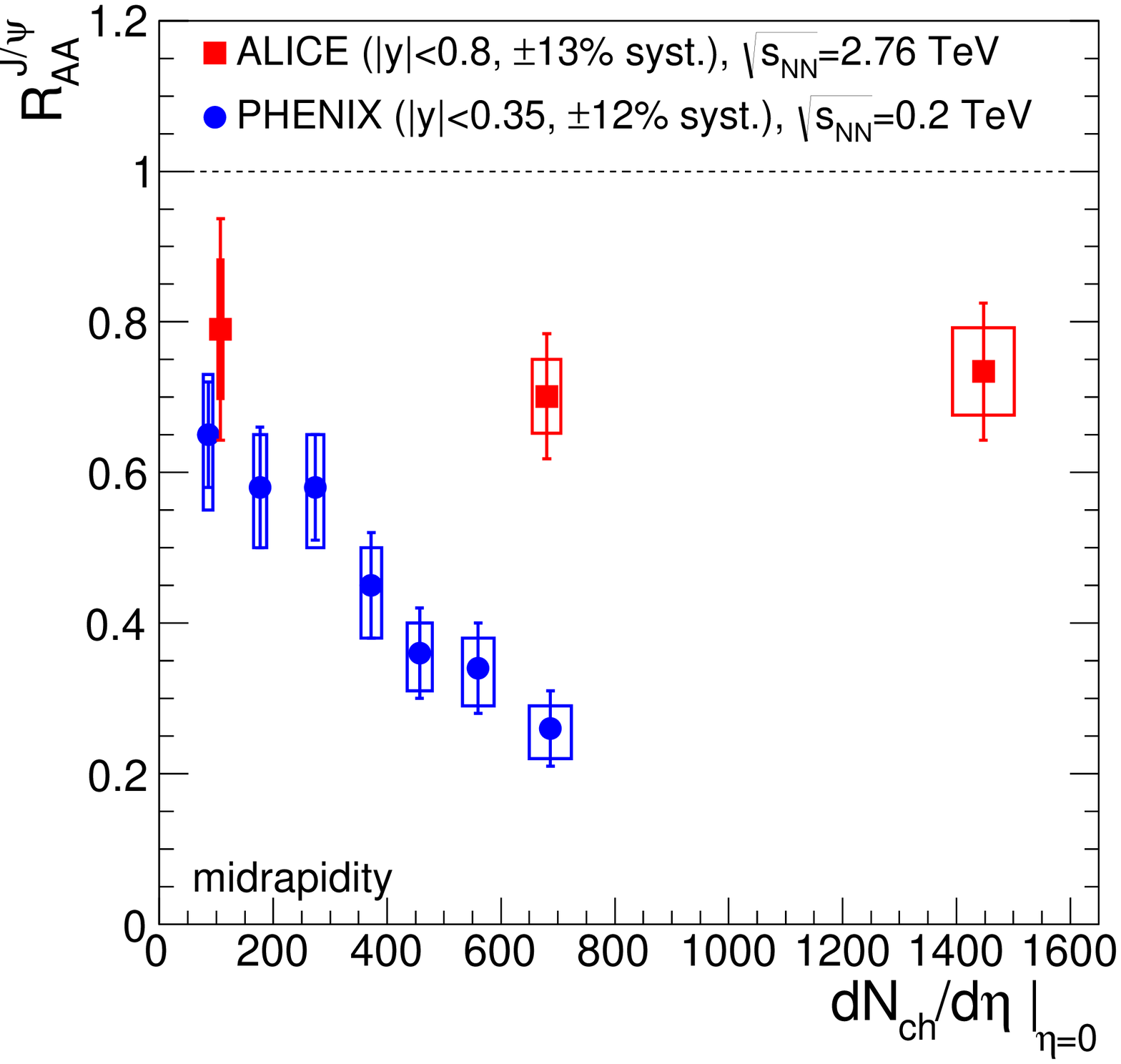}
\end{minipage} & \begin{minipage}{.49\textwidth}
\hspace{-0.3cm}\includegraphics[width=1.02\textwidth]{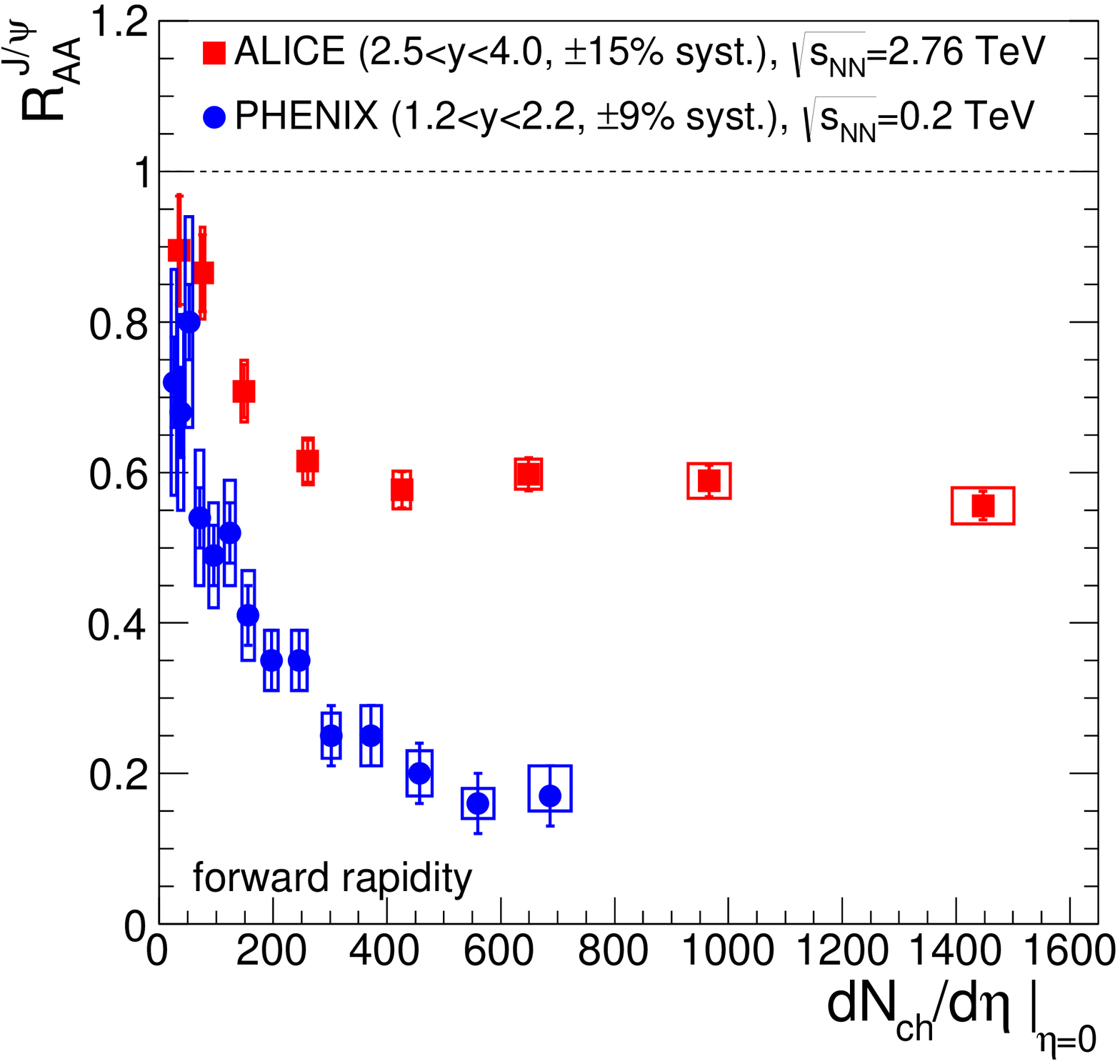}
\end{minipage}\end{tabular}
\caption{The dependence of the nuclear modification factor $R_{\mathrm{AA}}$ for
inclusive J/$\psi$ production on the multiplicity density (at $\eta$=0) 
for midrapidity (left panel) and at forward rapidity (right panel).
The data are integrated over $p_{\mathrm T}$  and are from the PHENIX collaboration
\cite{Adare:2006ns} at RHIC and ALICE collaboration \cite{Abelev:2013ila} at the LHC.
Note the overall uncertainties of the data quoted in the legend.
}
\label{fig:raa_jpsi1}
\end{figure}

The measurement of J/$\psi$ production in Pb--Pb collisions at the LHC was 
expected to provide a definitive answer on the question of (re)generation.
The data measured at high $p_{\mathrm T}$ \cite{Chatrchyan:2012np} show a 
pronounced suppression of J/$\psi$ in Pb--Pb compared to pp collisions
and of the same magnitude as that of open-charm hadrons, see 
Fig.~\ref{fig:raa_comp}.
This may indicate that the high-$p_{\mathrm T}$ charm quarks that form either 
$D$ or J/$\psi$ mesons had the same dynamics, possibly a thermalization 
in QGP and a late hadronization.

\begin{figure}[hbt]
\centering\includegraphics[width=.61\textwidth,height=.6\textwidth]{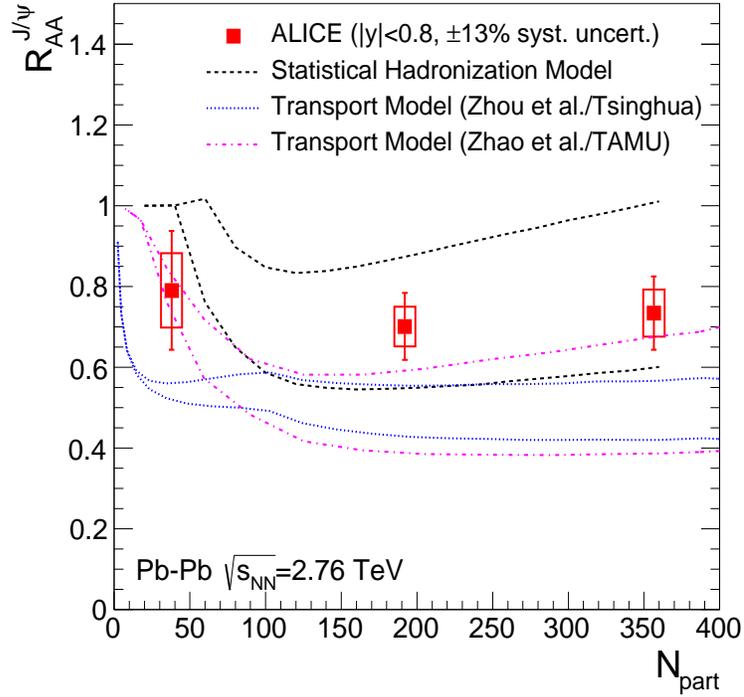}
\caption{Centrality dependence of the nuclear modification factor for inclusive 
J$/\psi$ production  at the LHC. The data at midrapidity \cite{Abelev:2013ila} 
are compared to model calculations: the statistical hadronization model 
\cite{Andronic:2011yq} and transport models of the TAMU \cite{Zhao:2011cv,Zhao:2012gc} 
and Tsinghua \cite{Liu:2009nb,Zhou:2014kka} groups. The bands are incorporating
(part of) the uncertainty in the charm production cross section.
} 
\label{fig:raa_jpsi2} 
\end{figure}

The first LHC measurement of the overall (inclusive in $p_{\mathrm T}$) 
production \cite{Abelev:2012rv}, showed for foward rapidities $R_{\mathrm{AA}}$ values 
significantly larger than at RHIC energies.
More recent data \cite{Abelev:2013ila} confirmed this, see Fig.~\ref{fig:raa_jpsi1}. 
The data are well described by both the statistical hadronization model 
\cite{Andronic:2011yq} and by transport models \cite{Liu:2009nb,Zhao:2011cv}, 
as demonstrated in Fig.~\ref{fig:raa_jpsi2}. 
Based on these observations, the J/$\psi$ production can be considered 
a probe of deconfinement as initially proposed \cite{Matsui:1986dk}, but may not be 
a  ``thermometer'' of the medium. 
Within the statistical model, the charmonium states become probes 
of the phase boundary between the deconfined and the hadron phases. 
This extends the family of quarks employed for the determination of 
the hadronization temperature (via the conjectured connection to the 
chemical freeze-out temperature extracted from fits of statistical model 
calculations to hadron abundances discussed above).

\begin{figure}[hbt]
\centering\includegraphics[width=.62\textwidth]{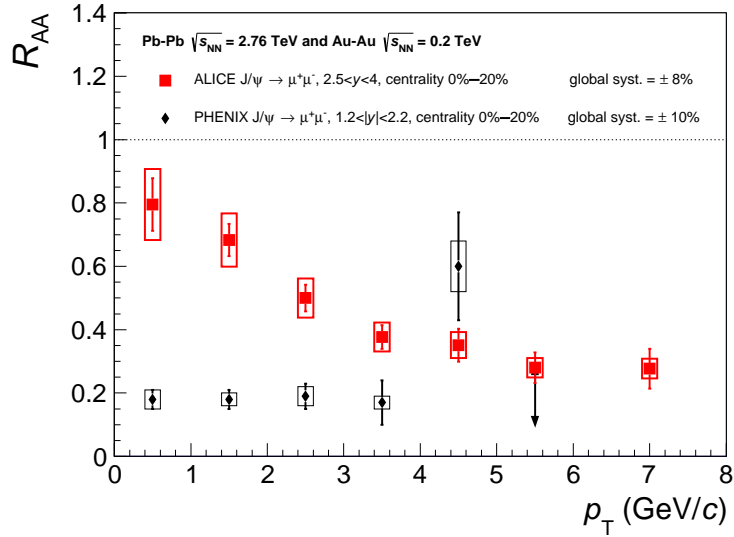}
\caption{Transverse momentum dependence of the nuclear modification factor 
for J/$\psi$ production measured at RHIC \cite{Adare:2006ns} and at
the LHC \cite{Abelev:2013ila} (plot from Ref. \cite{Abelev:2013ila}).
} 
\label{fig:raa_jpsi3} 
\end{figure}

In transport models, about 60-70\% of the J/$\psi$ yield in Pb--Pb collisions
originates from (re)combination of $c$ and $\bar{c}$ quarks, the rest
being primordial J/$\psi$ mesons that have survived in the deconfined medium.
These models show  \cite{Zhao:2011cv,Zhou:2014kka} that, as expected, (re)generation 
is predominantly a low-$p_{\mathrm T}$ phenomenon, as illustrated by the measured
$p_{\mathrm T}$ dependence of $R_{\mathrm{AA}}$, shown in Fig.~\ref{fig:raa_jpsi3}.
The measurement of J/$\psi$ elliptic flow at the LHC \cite{ALICE:2013xna}, albeit
to date of limited statistical significance, brings another argument in favor of
charm quark thermalization.
The J/$\psi$ data at RHIC are compatible with a null flow signal \cite{Adamczyk:2012pw}.
A $v_2$ signal was measured for J/$\psi$  at the SPS \cite{Prino:2008zz} and was 
interpreted as a path-length dependence of the screening.

The quarkonium data at the LHC demonstrate that (re)generation in deconfined matter 
during the QGP lifetime or at the chemical freeze-out are the only possible mechanisms 
of production. Discriminating the two pictures will help providing an answer to 
fundamental questions related to the fate of (quarkonium) hadrons in a hot 
medium \cite{Liu:2006nn,Mocsy:2007yj,Laine:2011xr}.
Data at the top LHC energy, including measurements on $\psi(2S)$ production, 
will help clarifying such questions.
Recent measurements of charmonium production in p-A collisions
at the LHC \cite{Abelev:2013yxa,Aaij:2013zxa} could help constraining models.
Interesting aspects are revealed by the measurement of $\psi(2S)$ production 
in d--Au collisions at RHIC \cite{Adare:2013ezl} and in p--Pb at the LHC \cite{Abelev:2014zpa}, indicating possible final-state effects.

\begin{figure}[hbt]
\centerline{\includegraphics[width=.6\textwidth]{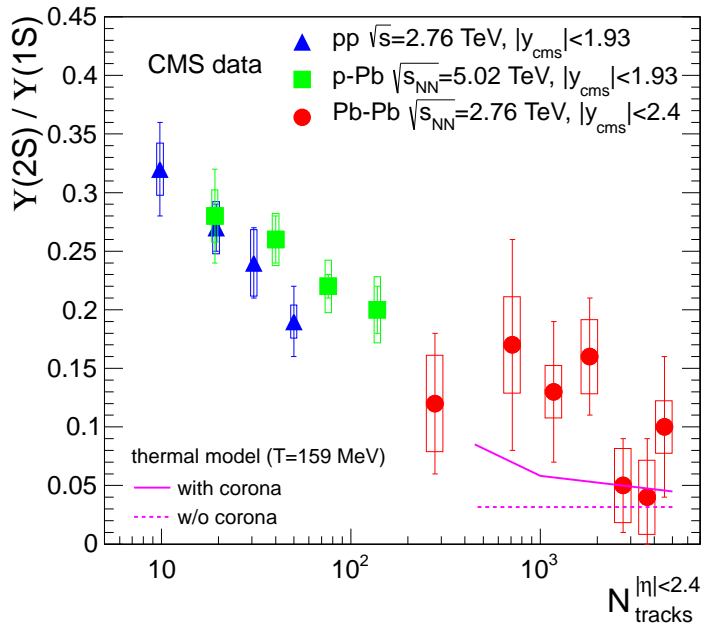}}
\caption{Track multiplicity dependence of the ratio of yields if $\Upsilon(2S)$ and
$\Upsilon(1S)$ bottomonium states, measured at the LHC in pp, p--Pb and 
Pb--Pb collisions \cite{Chatrchyan:2013nza}. 
The lines are thermal model predictions for central Pb--Pb collisions; 
the full line includes an estimate of the contribution of the production in the 
corona of the colliding nuclei.
} 
\label{fig:y} 
\end{figure}

Recent measurements of production of bottomonium ($b\bar{b}$) states at the 
LHC \cite{Chatrchyan:2012fr,Chatrchyan:2013nza,Abelev:2014nua} and at 
RHIC \cite{Adare:2014hje} add an important new aspect to the 
quarkonium dissociation story.
The nuclear modification factor for the $\Upsilon$ states exhibits at the LHC 
a sequential suppression pattern \cite{Chatrchyan:2012fr}. 
Transport model predictions \cite{Emerick:2011xu} describe the data (albeit not in 
detail at the LHC \cite{Abelev:2014nua}) while the Debye screening picture implemented 
in a hydrodynamical approach \cite{Strickland:2012cq} is less successful 
(see Ref. \cite{Abelev:2014nua}).
The production ratio $\Upsilon(2S)/\Upsilon(1S)$, shown in Fig.~\ref{fig:y},
is very different in Pb--Pb compared to pp collisions \cite{Andronic:2009sv};
the data are compatible, for central Pb--Pb collisions, to the value predicted by 
the statistical hadronization model for $T=159$ MeV.
This provides a tantalizing possibility of adding the bottom flavor towards
constraining even further the QCD phase boundary with nucleus-nucleus data at 
high energies.


\bibliography{NNrev}

\end{document}